\documentclass[]{spie}  %>>> use for US letter paper
\usepackage{amsmath,amsfonts,amssymb}
\usepackage{graphicx}
\usepackage{subfig}
\usepackage[colorlinks=true, allcolors=blue]{hyperref}

\title{Testing magnetic interference between TES detectors and the telescope environment for future CMB satellite missions}

\author[a]{Tommaso~Ghigna}
\author[a]{Thuong~D.~Hoang}
\author[a]{Takashi~Hasebe}
\author[b]{Yurika~Hoshino}
\author[a]{Nobuhiko~Katayama}
\author[c]{Kunimoto~Komatsu}
\author[d]{Adrian~Lee}
\author[a]{Tomotake~Matsumura}
\author[c]{Yuki~Sakurai}
\author[b]{Shinya~Sugiyama}
\author[e]{Aritoki~Suzuki}
\author[d]{Christopher~Raum}
\author[a]{Ryota~Takaku}
\author[d]{Benjamin~Westbrook}
\author[ ]{for~the~\textit{LiteBIRD}~Collaboration}
\affil[a]{Kavli IPMU (WPI), UTIAS, The University of Tokyo, Kashiwa, Chiba 277-8583, Japan}
\affil[b]{Saitama University, 255 Shimookubo, Sakura-ku, Saitama, 338-8570, Japan}
\affil[c]{Okayama University,  3-1-1, Tsushimanaka, Kita-ku, Okayama, 700-8530, Japan}
\affil[d]{Department of Physics, University of California Berkeley, Berkeley, 94720, California, USA}
\affil[e]{Lawrence Berkeley National Laboratory, 1 Cyclotron Rd, Berkeley, CA 94720, USA}

\authorinfo{Further author information: (Send correspondence to T.G.) E-mail: tommaso.ghigna@ipmu.jp}

\begin{document} 
\maketitle

\begin{abstract}
The two most common components of several upcoming CMB experiments are large arrays of superconductive TES (Transition-Edge Sensor) detectors and polarization modulator units, e.g. continuously-rotating Half-Wave Plates (HWP). A high detector count is necessary to increase the instrument raw sensitivity, however past experiments have shown that systematic effects are becoming one of the main limiting factors to reach the sensitivity required to detect primordial $B$-modes. Therefore, polarization modulators have become popular in recent years to mitigate several systematic effects. Polarization modulators based on HWP technologies require a rotating mechanism to spin the plate and modulate the incoming polarized signal. In order to minimize heat dissipation from the rotating mechanism, which is a stringent requirement particularly for a space mission like \textit{LiteBIRD}, we can employ a superconductive magnetic bearing to levitate the rotor and achieve contactless rotation. A disadvantage of this technique is the associated magnetic fields generated by those systems. In this paper we investigate the effects on a TES detector prototype and find no detectable $T_c$ variations due to an applied constant (DC) magnetic field, and a non-zero TES response to varying (AC) magnetic fields. We quantify a worst-case TES responsivity to the applied AC magnetic field of $\sim10^5$ pA/G, and give a preliminary interpretation of the pick-up mechanism.
\end{abstract}


\section{INTRODUCTION}
\label{sec:intro}  
The main scientific target of several upcoming CMB experiments \cite{simons_obs_2019, bicep_2022, litebird_ptep, cmbs4_2022} is the primordial CMB $B$-mode spectrum. If detected it will shed new light on the physics of the early Universe, in particular with regards to inflation \cite{guth_1981}. From current $B$-mode upper limits \cite{tristram_2022} we know that a detection is going to be challenging even for the most recent and advanced detector technologies. Hence, most upcoming and future experiments will observe the sky using multiple telescopes with focal planes made of several thousand superconductive Transition-Edge Sensor (TES) bolometers \cite{irwin_2005} to achieve the sensitivity required for a successful detection of this faint signal\cite{barron_2018}.

Unfortunately intrinsic detector noise is not the only limiting factor for a successful $B$-mode detection. Atmospheric fluctuations\cite{takakura_2019} (for ground-based experiments), Galactic foregrounds\cite{puglisi_2022} and instrumental systematic effects\cite{ghigna_2020, puglisi_2021 ,fabbian_2021, giardiello_2022, krachmalnicoff_2022} can contaminate the data and have a negative impact on the  experiment sensitivity.

A rapid modulation of the polarized sky signal has become a popular strategy to address and overcome some of these challenges. Polarization modulation can be achieved with a Polarization Modulation Unit (PMU) in front of the focal plane, which in most cases consists of a Half-Wave Plate (HWP) rotated at constant speed. This technique offers several advantages. Since the signal of interest from the sky is polarized, by appropriately tuning the HWP rotation speed, it is possible to modulate the signal in a frequency range free from low-frequency atmospheric and instrumental $1/f$ noise\cite{takakura_2017}. If the HWP is the first optical element in
\begin{figure}[htbp]
    \centering
    \includegraphics[width=1\textwidth]{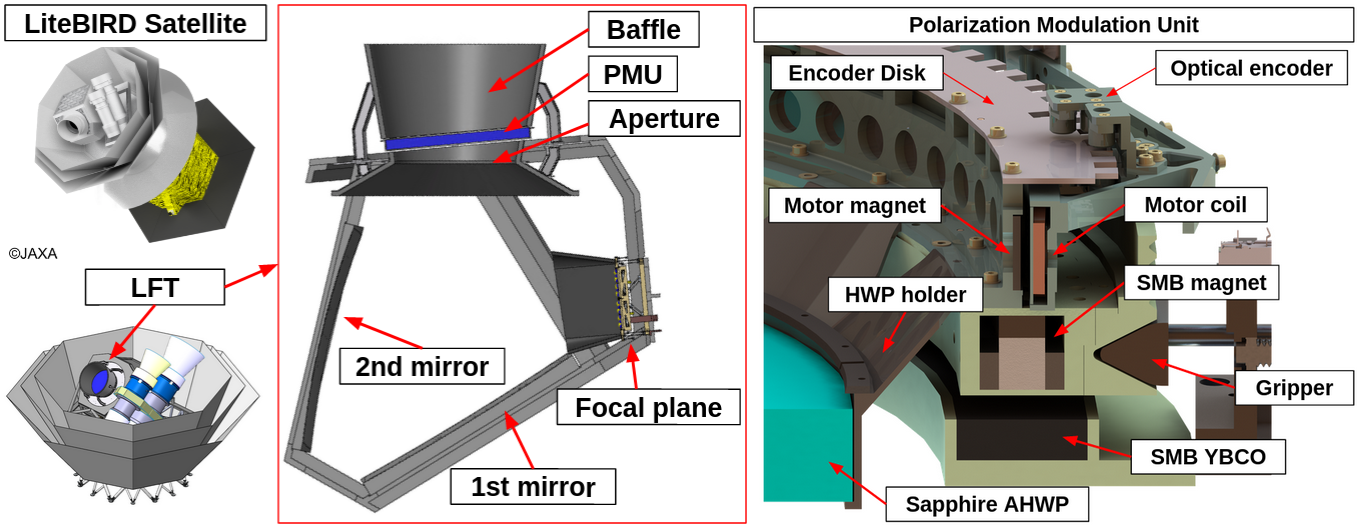}
    \caption{Left: A snapshot of \textit{LiteBIRD} satellite and payload with the three telescopes for low-frequency (LFT), middle-frequency (MFT) and high-frequency (HFT) detectors \cite{litebird_ptep}. Center: A cross-sectional view of LFT where the PMU and superconductive focal plane are highlighted\cite{sekimoto_2020}. Right: LFT Polarization Modulation Unit design and components. Notice the presence of multiple permanent magnets in the SMB and the driving motor\cite{sakurai_2020}.}%
    \label{fig:litebird_pmu}
\end{figure}
the telescope system, instrumental polarization effects are not modulated and can be easily separated from the sky signal. Modulation of the polarized signal allows also for the reconstruction of the polarization Stokes $Q$ and $U$ parameters with a single detector, reducing the impact of differential systematic effects due to intrinsic differences in a pair of polarization sensitive detectors\cite{hoang_2017}.

In order to minimize unwanted HWP non-ideal effects it is necessary to operate the HWP at a stable cryogenic temperature to avoid temperature fluctuations and reduce the HWP emissivity\cite{takakura_2017, takaku_2022}. Opting for this choice requires placing the PMU inside the receiver dewar. This imposes tight constraints on the heat dissipation of the rotating mechanism. For the \textit{LiteBIRD} space mission Low Frequency Telescope the thermal budget allocated to the PMU is 4 mW\cite{hasebe_2022}. For this reason we are developing a contactless rotation mechanism for LFT PMU. The prototype consists of a Superconductive Magnetic Bearing (SMB) to levitate the rotor that supports the HWP, and a magnetic motor to spin the rotor at constant speed\cite{sakurai_2020}. In Figure \ref{fig:litebird_pmu} we show a snapshot of the \textit{LiteBIRD} satellite, payload, a cross-sectional view of LFT and the most recent PMU design.

The unfortunate byproduct of these components is a non-zero magnetic field at the detector location on the focal plane (if not properly shielded). For \textit{LiteBIRD}, a detailed analysis of the magnetic field strength (and its possible time-dependence given the use of a magnetic motor made of moving segmented permanent magnets) is still being investigated. In preparation, we have begun addressing the impact of magnetic fields on \textit{LiteBIRD} TES detector prototypes using a Helmholtz coil to generate a stable magnetic field.

In recent years some measurements of magnetic field dependent TES behaviour have been reported, focusing in particular on the observed magnetic field dependence of the critical temperature $T_c$ of AlMn superconductive films\cite{vavagiakis_2018}. \textit{LiteBIRD} will employ similar Mn-doped aluminium TES bolometers. Our final goal is to determine the impact of magnetic fields generated by components of the \textit{LiteBIRD} telescope (e.g. the PMU) on \textit{LiteBIRD}-like superconductive bolometers to define the best magnetic shield configuration around the focal plane. In this paper 
\begin{table}[htbp]
    \centering
    \begin{tabular}{c c c c c c}
        Leg length [$\mu$m] & Leg cross-section [$\mu$m] & AlMn area [$\mu$m$^2$] & AlMn thickness [nm] & $T_c$ [K] & $R_{\rm n}$ [$\Omega$]\\
        \hline
        500 & 1$\times$17 & 480 & 90 & 0.220 & $0.78$ \\
    \end{tabular}
    \caption{Size and critical temperature of the TES detector tested for this work for comparison with Vavagiakis et al\cite{vavagiakis_2018}. The AlMn film is made with aluminum doped with 5500 ppm of manganese by weight. AlMn size and normal resistance are in the same range as POLARBEAR-like detectors shown in Vavagiakis et al., which also appear to show low $T_{\rm c}$ sensitivity to an applied DC magnetic field.}
    \label{tab:detector}
\end{table}
\begin{figure}[htbp]
    \centering
    \subfloat{{\includegraphics[width=.554\textwidth]{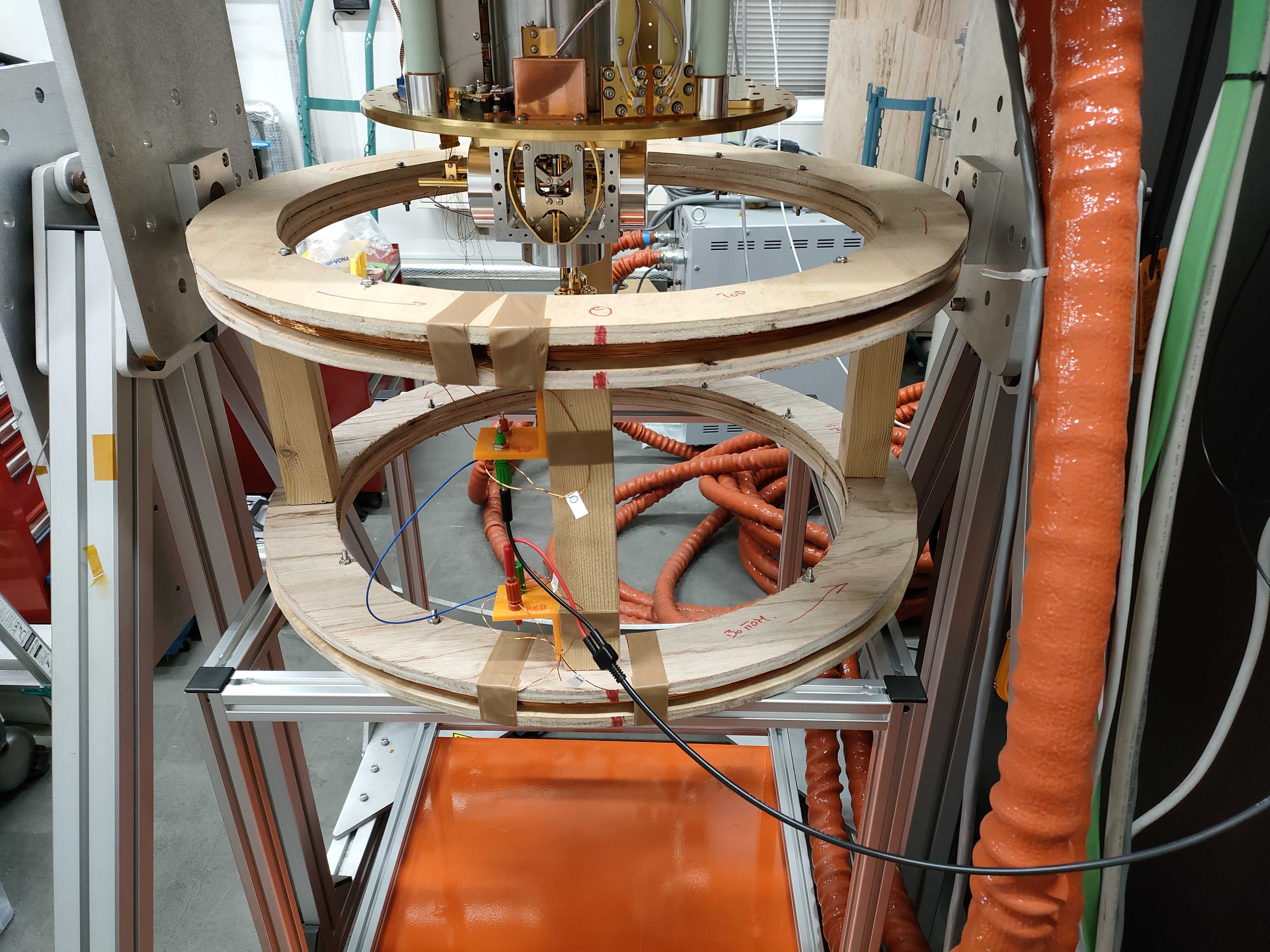} }}%
    \subfloat{{\includegraphics[width=.446\textwidth]{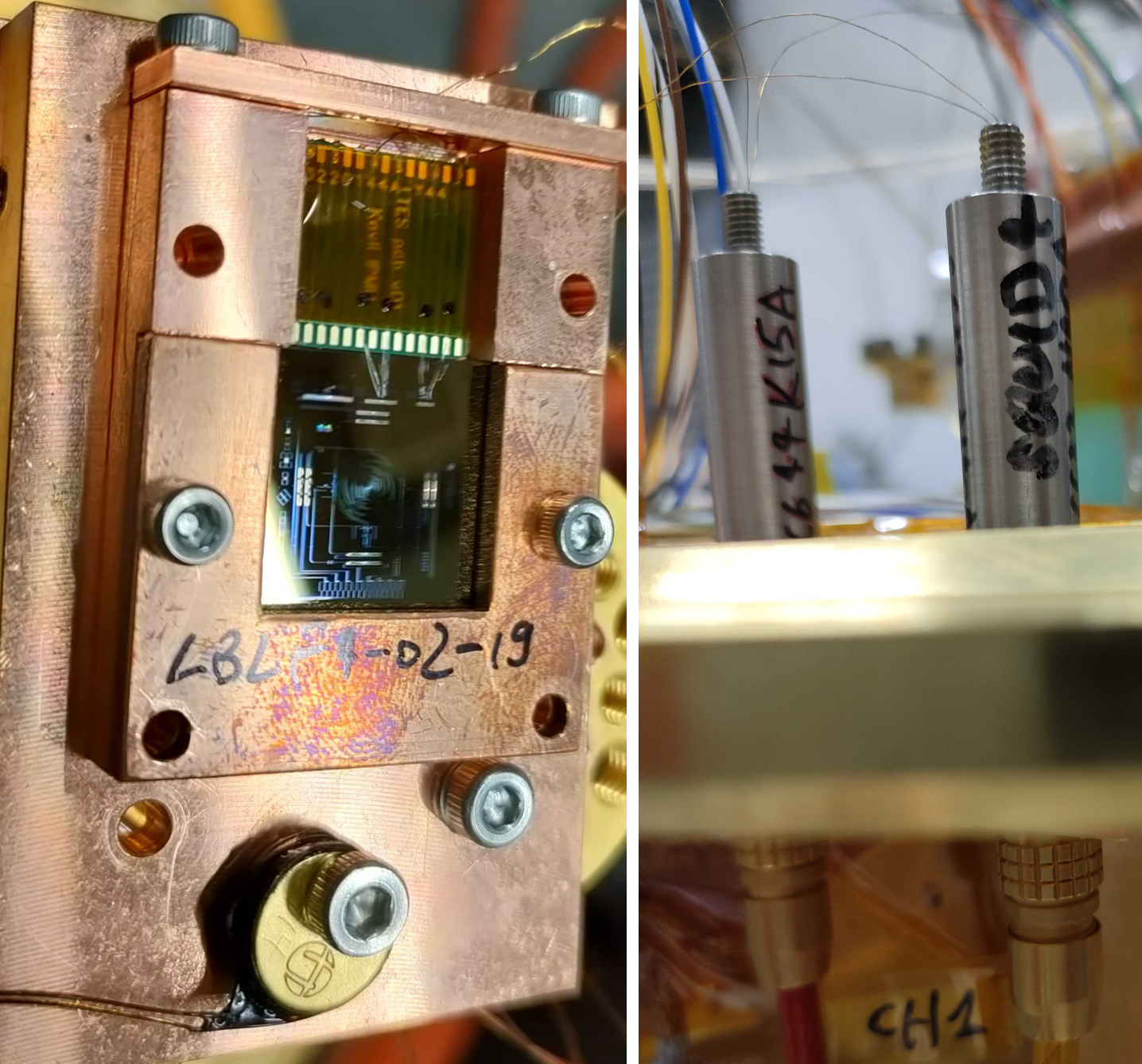} }}%
    \caption{Left: the Helmholtz coil positioned around the open ADR cryostat at Kavli IPMU. Center: the TES chip tested for this work. Right: two SQUID amplifiers with their Nb can shield.}%
    \label{fig:helmholtz}
\end{figure}
we present the system we have developed for an initial assessment of the magnetic field sensitivity of an AlMn TES prototype. The prototype tested for this paper came from an early fabrication run and does not fully meet \textit{LiteBIRD} specifications, in particular we measured $T_c\sim 220$~mK and $P_{\rm sat}\sim4$~pW, while $T_c\sim 170$~mK and $P_{\rm sat}\sim1$~pW are the target\cite{westbrook_2020, westbrook_2022} (more design and measured parameters of the TES tested are reported in Table \ref{tab:detector}). Nevertheless, we decided to report these results given the scarcity of data about TES magnetic field sensitivity in the literature.

\section{DC MAGNETIC FIELD}
\label{sec:dc_magnetic_field}
Magnetic field sensitivity of AlMn TES critical temperature has been reported in Vavagiakis et al.\cite{vavagiakis_2018}. The authors tested several different TES detectors (from ACT \cite{niemack_2010, thornton_2016} and POLARBEAR \cite{inoue_2016} experiments) of different size (AlMn area, thickness and thermal link), normal resistance $R_{\rm n}$ and nominal critical temperature $T_c$. For most of the detector tested $T_c$ appears to be decreasing with the strength of the applied magnetic field.
In particular, the detectors of lower normal resistance $R_{\rm n}$ ($R_{\rm n}\sim0.007$ $\Omega$ for ACT-like detectors) appear to be more sensitive to the applied magnetic field compared to those of higher resistance ($R_{\rm n}\sim0.8$ $\Omega$ for POLARBEAR-like detectors).

Testing such dependence is important in the design phase of a CMB experiment because a change in $T_c$ can impact two fundamental detector parameters: saturation power $P_{\rm sat}$ and the phonon noise $NEP_{G}$ that originates in the thermally conductive legs that keep the absorber-thermistor island isolated from the thermal bath (this is the dominant contributor to the detector intrinsic noise). These two parameters can be obtained knowing the temperature of the thermal bath $T_b$ and the detector critical temperature $T_c$. If we assume the thermal conductivity of the legs to be of the form $k_{\rm 0}T^{n}$ (where $k_{\rm 0}$ is a constant and $n$ is the carrier index that defines the temperature dependence), we can write\cite{suzuki_phd}:
\begin{equation}
    \label{eq:psat}
    P_{\rm sat} = N \frac{A}{\ell}\frac{k_{\rm 0}}{n+1}(T_c^{n+1}-T_b^{n+1}),
\end{equation}
\begin{equation}
    \label{eq:nep}
    NEP^2_{G} = 4P_{\rm sat}k_{b}T_{b}\frac{(n+1)^2}{2n+3}\frac{(T_c/T_b)^{2n+3}-1}{[(T_c/T_b)^{n+1}-1]^2},
\end{equation}
where $N$, $A$ and $\ell$ are respectively the number, cross-section area and length of the thermally conductive legs, and $n$ is the thermal conductivity index. For purely phonon-mediated heat transfer the index is assumed to be $n=3$. However values in the range $1<n<3$ have been measured for different detector materials and designs\cite{jaehnig_2022}. 
\begin{figure}[htbp]
    \centering
    \subfloat{{\includegraphics[width=.5\textwidth]{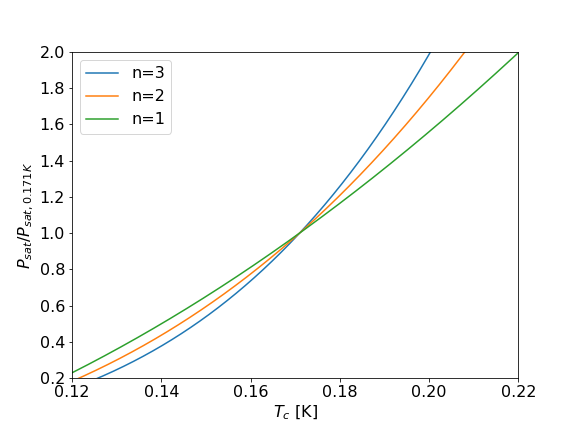} }}%
    \subfloat{{\includegraphics[width=.5\textwidth]{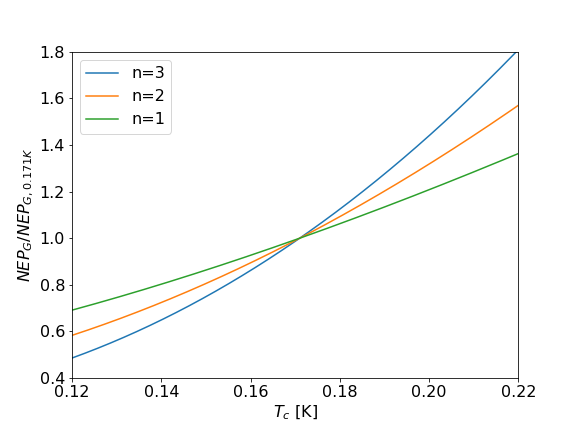} }}%
    \caption{Left: Saturation power $P_{\rm sat}$ vs. bolometer transition temperature $T_c$ for different values of the thermal conductivity index $n$ assuming bath temperature $T_b=0.1$ K. Right: Phonon Noise Equivalent Power $NEP_{G}$ vs. bolometer transition temperature $T_c$ for different values of the thermal conductance index $n$ assuming bath temperature $T_b=0.1$ K and fixed $P_{\rm sat}$. In both cases the curve is normalised to the expected value for $T_c=0.171$ K (baseline value for \textit{LiteBIRD} detectors).}%
    \label{fig:nep_psat}
\end{figure}

Assuming \textit{LiteBIRD} baseline values for $T_b=100$ mK and $T_c=171$ mK, we have computed $P_{\rm sat}$ and $NEP_G$ (for fixed $P_{\rm sat}$ value) as a function of $T_c$ for three different $n$ values. In Figure \ref{fig:nep_psat} we show the results normalized to the expected value for $T_c=171$ mK. It is clear from Figure \ref{fig:nep_psat} that a change of $T_c$ can have an impact on the effective value of these two parameters. Specifically, a $T_c$ value lower than expected will result in a reduction of both $P_{\rm sat}$ and $NEP_G$. While a better noise performance can potentially be beneficial, a reduction of $P_{\rm sat}$ changes the detectors dynamic range, and could cause them to saturate more easily when observing a bright source.

For this reason, we need to understand the effect of magnetic fields on the critical temperature of \textit{LiteBIRD} detectors to design the most appropriate magnetic shield to avoid $T_c$
\begin{figure}[htbp]
    \centering
    \subfloat{{\includegraphics[width=.5\textwidth]{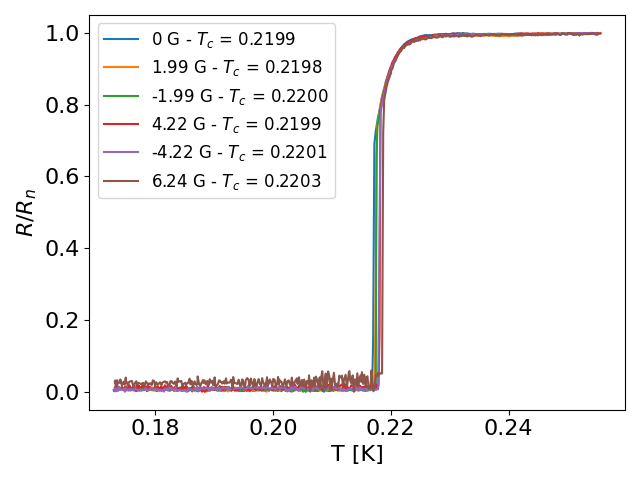} }}%
    \subfloat{{\includegraphics[width=.5\textwidth]{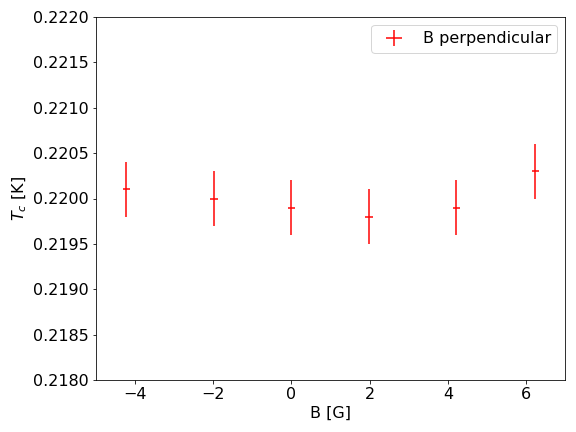} }}%
    \caption{Left: Temperature vs. resistance of the TES tested for this work. The measurement has been repeated for different applied constant magnetic fields perpendicular to the TES chip plane. Right: Critical temperature measured from the plot on the left at $R=0.9\times R_{\rm n}$. The errorbar corresponds to the measured temperature accuracy $\sigma_{\rm T}=0.0003$ K.}%
    \label{fig:T_vs_R}
\end{figure}
\begin{figure}[htbp]
    \centering
    \subfloat{{\includegraphics[width=.5\textwidth]{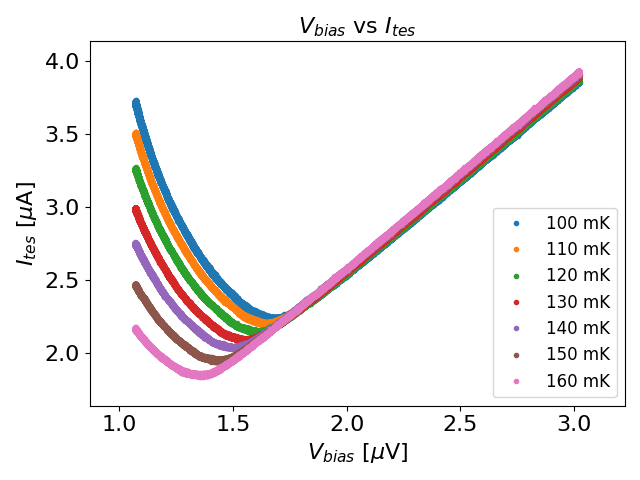} }}%
    \subfloat{{\includegraphics[width=.5\textwidth]{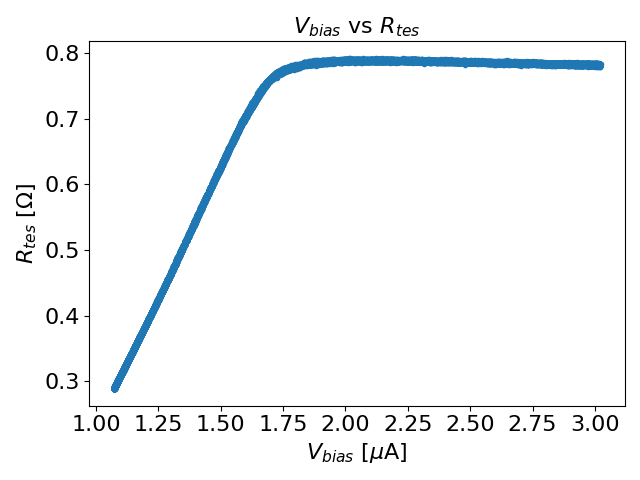} }}%
    \caption{Left: Voltage bias vs. TES current for different bath temperatures $T_b$. For the magnetic field sensitivity measurements we fixed $T_b$ to 100 mK. Right: Voltage bias vs. TES resistance.}%
    \label{fig:tes_characterization}
\end{figure}
variations that could potentially alter detector properties. For this reason we have assembled a Helmholtz coil to generate a magnetic field to test detector prototypes.

In Sugiyama et al.\cite{sugiyama_2022}, we reported the expected value of the DC magnetic field produced by the PMU prototype at \textit{LiteBIRD} LFT focal plane: $B\sim0.24$ G. Since this is only a forecast and there may be other magnetic field sources in the fully-assembled system, we tested the detector under the effect of larger magnetic fields. In Figure \ref{fig:T_vs_R} we show the resistance-temperature relation measured for different applied magnetic field values in the range $-4.22$ G $<~B~<$ $6.24$ G, for the detector in Table \ref{tab:detector}. The magnetic field was applied perpendicularly to the AlMn surface. The $T_c$ value reported for each curve corresponds to the temperature at $R$~=~$0.9\times R_{\rm n}$. In the right panel we also show the $T_c$ value extrapolated from the data in the left panel. We do not detect any significant variation of $T_c$ within the measured temperature accuracy.

While no significant shift in the resistance-temperature relation is visible around the turnaround at $R\sim0.9\times R_{\rm n}$, a small shift can be observed deeper into the superconductive transition. This appears to be an artifact attributed to a noise increase for stronger magnetic fields (each measurement was repeated twice to test the reproducibility of the results). 

\section{AC MAGNETIC FIELD}
\label{sec:ac_magnetic_field}
Given the presence of active components, moving parts and even moving permanent magnets\cite{sakurai_2020} in the telescope system of upcoming CMB experiments (the PMU employ a magnetic motor with segmented permanent magnets in the rotor), we can expect time-dependent magnetic fields to be generated. While a full assessment of the origin and magnitude at the focal plane has yet to be evaluated for the \textit{LiteBIRD} experiment, we have measured the displacement of the rotor of the PMU prototype for \textit{LiteBIRD} LFT during contactless rotation of the rotor. In Sugiyama et al.\cite{sugiyama_2022} we presented the measurement and a forecast of the expected AC magnetic field amplitude at the focal plane location due to the rotor displacement: $B_{\rm m}\sim 3\times10^{-5}$ G. While at present such a small magnetic field is challenging to reliably generate and measure, we have used the Helmholtz coil in combination with a commercial signal generator to test the TES succeptibility to an AC magnetic field.
\begin{table}[htbp]
    \centering
    \begin{tabular}{c||c c c c c c}
        $V_b$ [$\mu$V] & 2.0 & 1.6 & 1.4 & 1.2 & 1.1 & 1.0 \\
        \hline
        $\mathcal{L}$ & 0.32 & 2.02 & 3.04 & 6.54 & 10.95 & 27.72 \\
    \end{tabular}
    \caption{Loop-gain values estimated from time-constant measurement: $\tau_{\rm eff}=\tau_{\rm 0}/(\mathcal{L}+1)$. The normal time-constant $\tau_0$ has been estimated as $\sim 13.5$ ms. The effective time-constant is estimated from bias-step measurements.}
    \label{tab:loop_gain}
\end{table}
\begin{figure}[htbp]
    \centering
    \subfloat{{\includegraphics[width=.495\textwidth]{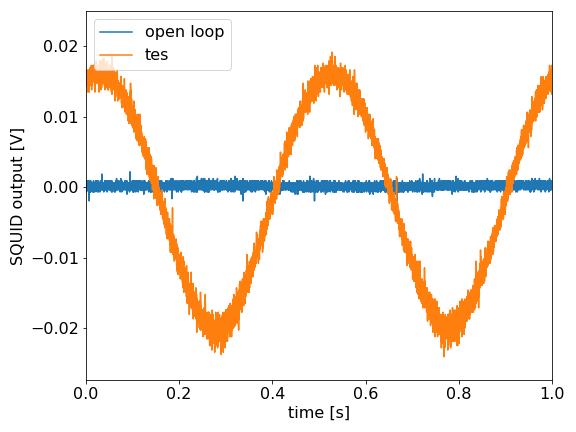} }}%
    \subfloat{{\includegraphics[width=.505\textwidth]{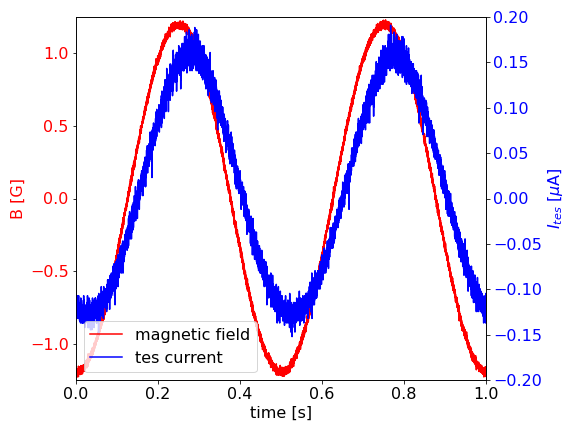} }}%
    \caption{Left: AC magnetic field pick-up. The blue line shows the output for a magnetic field shielded SQUID with open input loop. The orange line shows the output when a TES biased in the superconductive transition is connected to the SQUID input loop. In both cases the magnetic field amplitude is $B_{\rm m}\sim1.4$ G and is modulated at 2 Hz. Right: An example of the measurement performed for this analysis. The red line shows the applied AC magnetic field, while the blue line is the TES current due to the magnetic field pickup.}%
    \label{fig:mag_field_pick_up}
\end{figure}

Since we expect the succeptibility of the TES detector to depend on the exact bias conditions during the test, in Figure \ref{fig:tes_characterization} we report a measurement of the voltage-current relation (V-I curve) for different values of the thermal bath $T_b$, as well as the voltage-resistance relation for $T_b=100$~mK. In Table \ref{tab:loop_gain} we also report an estimate of the loop-gain $\mathcal{L}$\cite{irwin_2005} for different voltage bias values determined from time-constant measurements performed with a bias-step analysis\cite{ghigna_jltp_2020}. These values are obtained by biasing the TES at voltage $V_b$ in the range 1-2~$\mu$V and then injecting a voltage step of 100~nV with a square-wave generator. The decaying tail is fitted with an exponential function to estimate the thermal time constant $\tau_{\rm eff}$. We assumed the normal time constant to be $\tau_0\sim13.5$~ms as measured above the TES transition at $V_b=2.5$~$\mu$V. Assuming this as the true value we can estimate the loop-gain $\mathcal{L}$ for all bias cases from the relation $\tau_{\rm eff}=\tau_{\rm 0}/(\mathcal{L}+1)$.
\begin{figure}[htbp]
    \centering
    \subfloat{{\includegraphics[width=.5\textwidth]{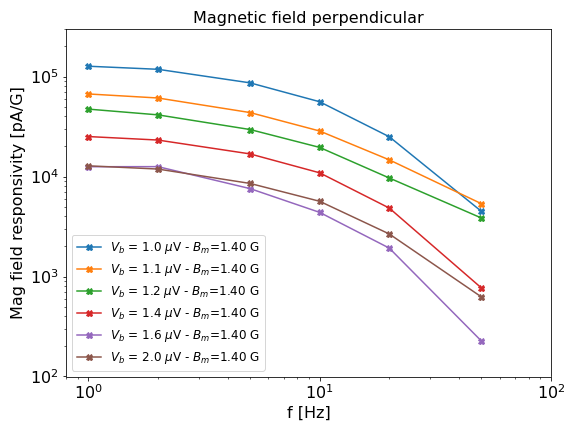} }}%
    \subfloat{{\includegraphics[width=.5\textwidth]{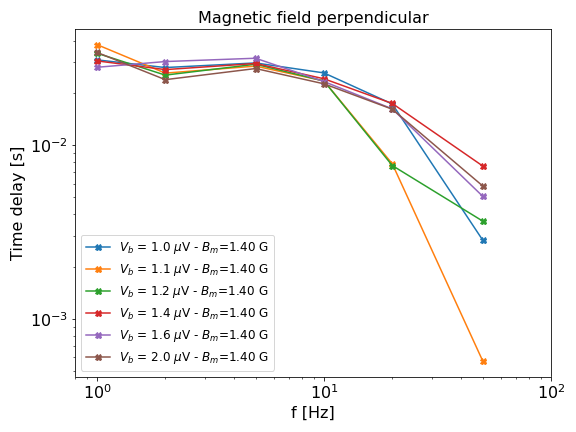} }}%
    \caption{Left: TES AC-magnetic field sensitivity for different bias conditions and applied field frequency. In all cases the amplitude of the magnetic field applied is $B_{\rm m}$=1.40 G. Right: Measured time-delay between the applied AC-magnetic field and the TES response for different bias voltage conditions $V_b$.}%
    \label{fig:tes_mag_field_sensitivity}
\end{figure}
\begin{figure}[htbp]
    \centering
    {\includegraphics[width=1\textwidth]{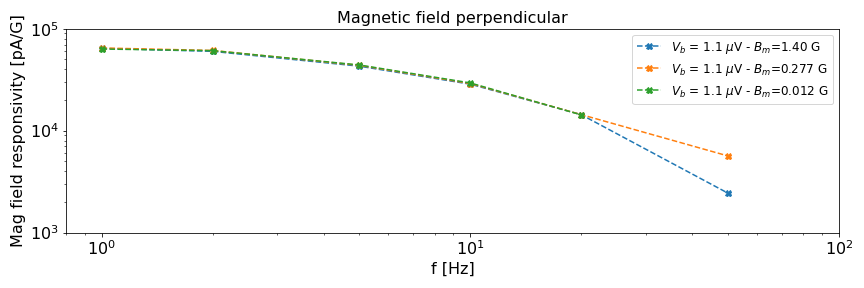} }
    \caption{Left: TES AC-magnetic field sensitivity for different amplitude of the applied magnetic field $B_{\rm m}$ under the same bias condition $V_b$=1.1 $\mu$V.}%
    \label{fig:tes_mag_field_sensitivity_amplitude}
\end{figure}

We read the TES with a commercial dc-SQUID amplifier in a Flux-Locked Loop (FLL) readout scheme\cite{ghigna_phd}. All SQUID amplifiers are magnetically shielded in a niobium can (shown in Figure \ref{fig:helmholtz}). Since a non-shielded SQUID could directly pick-up the magnetic field produced by the Helmholtz coil, in the left plot of Figure \ref{fig:mag_field_pick_up} we show the SQUID output voltage measured for the same SQUID amplifier in two different configurations: without the TES connected at the SQUID input (open loop) and with the TES under test connected to the SQUID input.
In both cases a sinusoidal AC magnetic field of amplitude $B_{\rm m}=1.4$~G amplitude (at the detector location) and 2 Hz frequency is applied. We can observe that for the open loop measurement the SQUID output does not seem to be affected by the presence of the magnetic field, indicating that the magnetic field pick-up is happening through the TES connected to the SQUID input.

As an example of the type of analysis we have performed, in the right plot of Figure \ref{fig:mag_field_pick_up} we show in blue the TES current $I_{\rm tes}$ calculated from the SQUID output in the left panel (orange line) and the magnetic field produced by the Helmholtz coil at the TES location. 

We tested the detector response for all six bias cases in Table \ref{tab:loop_gain} for different magnetic field amplitudes $B_{\rm m}$ and frequencies in the range 0.01-1.4 G and 1-50 Hz. Smaller and slower fields could not be generated because of limitations of the signal generator used. We were also limited by the readout noise. In order to increase the signal-to-noise ratio, we had to integrate the data over several magnetic field periods (in particular for small field amplitudes and high frequency). While in a realistic scenario we expect smaller fields than the ones tested, this analysis gives us a first assessment of the TES responsivity to an AC magnetic field. We plan to explore the lower frequency regime in the future, however the frequency range 1-10 Hz is particularly relevant in the CMB context because both \textit{LiteBIRD} and SO PMUs will be operated in this regime.   

From the data we determined the TES responsivity to the applied magnetic field by fitting the TES current and magnetic field data in the right panel of Figure \ref{fig:mag_field_pick_up} with a function of the form $A\sin{2\pi f (t-t_{\rm 0})}$. Where $A$ is the amplitude, $f$ is the frequency and $t_{\rm 0}$ is a time offset. We computed the responsivity to the applied magnetic field from the fit parameters as $A_{\rm I}/A_{\rm B}$ where $A_{\rm I}$ is the measured TES current amplitude and $A_{\rm B}$ is the magnetic field amplitude. We also measured the time-constant as the time-delay between the applied magnetic field and the detector response $|t_{\rm 0B}-t_{\rm 0I}|$.

In the left panel of Figure \ref{fig:tes_mag_field_sensitivity} we show the responsivity results $A_{\rm I}/A_{\rm B}$ for all bias cases in Table \ref{tab:loop_gain} for an applied magnetic field of amplitude $B_{\rm m}=1.4$ G perpendicular to the TES chip surface and for different frequencies. We can notice that the responsivity increases as the detector is biased deeper into the superconductive transition, while it decreases at higher frequencies. The responsivity appears to increase by an order of magnitude consistently at all frequencies with an increase of the loop-gain by roughly the same magnitude.

In the right panel of Figure \ref{fig:tes_mag_field_sensitivity} we show the estimated time-constant which appears to be consistently $\sim 30$~ms for most frequencies and bias conditions. A drop is observed for $f>20$ Hz.
\begin{figure}[htbp]
    \centering
    \includegraphics[width=1\textwidth]{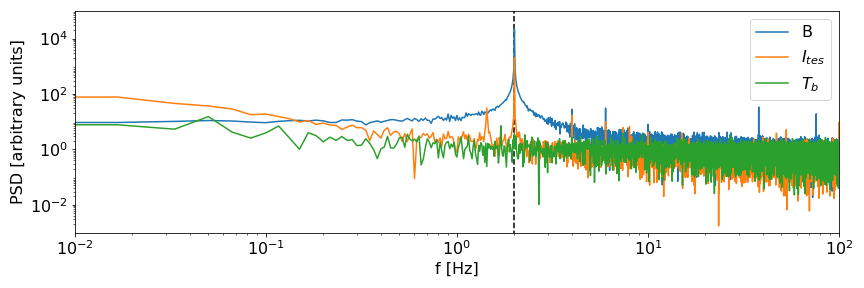}
    \caption{The PSD of the magnetic field $B$, TES current $I_{\rm tes}$ and a thermometer placed next to the TES chip at 100 mK to monitor the bath temperature $T_b$. The data correspond to Figure \ref{fig:mag_field_pick_up}. We do not observe variations of the bath temperature $T_b$ correlated to the 2 Hz AC magnetic field. Hence we can exclude eddy currents produced in the TES chip copper holder by the changing magnetic field to be responsible for bath temperature variations. The PSDs have been normalized to the white noise level for clarity.}
    \label{fig:temperature_stability}
\end{figure}
This is probably just an artifact due to  the magnetic field frequency approaching the inverse of the time-constant.  

In Figure \ref{fig:tes_mag_field_sensitivity_amplitude} we report the responsivity $A_{\rm I}/A_{\rm B}$ calculated from three different amplitudes of the applied magnetic field for the bias case $V_{\rm b}=1.1$ $\mu$V. This is a consistency check to confirm that the estimated responsivity does not change for different amplitudes of the applied field, which appears to be true at least for $f<20$ Hz. 
Given the smaller signal-to-noise ratio for high frequency fields, we attribute this discrepancy to the relative noise level increase.

Finally, in Figure \ref{fig:temperature_stability} we show the temperature stability of the bath temperature $T_b$ when an AC magnetic field of amplitude $B_{\rm m}=1.4$ G and frequency 2 Hz is applied. We show this data to exclude the possibility that the pick-up mechanism may be due to bath temperature fluctuations possibly originating from eddy currents produced by the magnetic field in the TES chip copper holder. It is in fact clear from the thermometer PSD that a 2 Hz component is not present.

Another possibility to explain the observed behaviour are eddy currents in the TES BLING (a metal layer that acts as a heat capacity to tune and slow down the TES thermal response) that can result in heat dissipation directly in the TES island. At present this remains an hypothesis, but we will try to address it in the future.

From preliminary results of the same analysis performed with a magnetic filed tangent to the TES AlMn area, we have initial indication that the responsivity could be orientation dependent. Unfortunately during these later tests we could not control accurately the bath temperature because of cryogenic issues, and the data appear to be noisy. Hence, we decided not to report these results here. However, we believe that testing the responsivity for different magnetic field orientations can help us understanding the pick-up mechanism, therefore we will certainly address this aspect in the future. 

\section{CONCLUSIONS}
\label{sec:conclusions}
We have reported on the DC and AC magnetic field sensitivity of TES detectors. In this work the magnetic field has always been applied perpendicularly to the TES AlMn area. In the future we will test the effect for different magnetic field orientations. We have found that, for the AlMn detector tested for this work, the critical temperature does not appear to be sensitive to an applied constant magnetic field in the range $-4.22$~G~$<B<6.24$~G. This result is in agreement previous results reported for AlMn detectors of similar normal resistance and physical size. This is a preliminary result that needs to be confirmed for future \textit{LiteBIRD} detector prototypes. 

On the other end, we have found the TES chip to be responsive to AC magnetic fields. We need more investigation to understand the pick-up mechanism. At present we have excluded the SQUID amplifier to be the source of the magnetic field pick-up as shown in the left panel of Figure \ref{fig:mag_field_pick_up}.

From the data presented here, we can not completely rule-out the possibility that the main pick-up mechanism is an inductive coupling between the TES biasing circuit (which is difficult to shield in the current configuration) and the varying magnetic field. More measurements of different TES detectors and for different magnetic field orientations may be needed to fully understand the mechanism.

However, we can speculate that if this was the main pick-up mechanism, the use of Digital Frequency-Domain Multiplexing (DfMux) readout system\cite{dobbs_2012} (like the one \textit{LiteBIRD} will employ) could potentially help mitigating the severity of this effect, given the presence of MHz LC filters (needed for multiplexing) in the TES biasing circuit. In fact, such filters can prevent unwanted low-frequency electrical signals from coupling to the TES response.

In Figure \ref{fig:tes_mag_field_sensitivity} we have shown how the responsivity appears to increase for low voltage bias values (smaller TES resistance and higher loop-gain). On the other end, the time-constant associated with the magnetic field pick-up does not change with bias conditions and appears to be stable around 30 ms. This value does not match the expected thermal or electrical time-constant of the detector-readout system. We will need more investigation to understand the pick-up mechanism.

As mentioned in Section \ref{sec:ac_magnetic_field}, we expect AC variations of the magnetic field generated by the rotating PMU of \textit{LiteBIRD} LFT with amplitude $B_{m}\sim3\times 10^{-5}$ G\cite{sugiyama_2022}. If we assume the worst case scenario responsivity (for a detector biased at $V_b=1\mbox{ }\mu$V) $A_{\rm I}/A_{\rm B}\sim10^5$ pA/G from Figure \ref{fig:tes_mag_field_sensitivity}, the current response for the expected magnetic field (assuming that this effect is linear as suggested by the results in Figure \ref{fig:tes_mag_field_sensitivity_amplitude}) would be $\Delta I_{\rm tes}\sim 3$ pA, for a non-shielded TES. This value is well below the current best readout noise level measured for TES-DfMux systems\cite{montgomery_2022, elleflot_2022} of NEI~$\sim 10\mbox{ pA}\sqrt{\mbox{Hz}}$ (noise equivalent current). While we can expect improvements with the noise performance of this technology in the coming years, we can speculate that such a small response is still going to be subdominant with respect to statistical noise sources, in particular when we take into account not only readout noise, but also the TES intrinsic noise and the dominant photon noise. A direct comparison of statistical noise sources with a systematic noise source is perhaps difficult and need to be clarified, given that the effect of the latter does not average down with longer integration time. Therefore, in the future we should focus not only on its amplitude but we should understand its time-evolution and how it could interact with the scan-strategy. It is likely that the signal is going to be synchronized with the HWP rotation. If so, we should understand which harmonic of the HWP rotation frequency is going to be dominant.

In conclusion, from the data presented here we expect this effect to be dominated by statistical noise, once an appropriate magnetic shielding is included to suppress the magnetic field amplitude. However, we believe that more investigation is necessary for lower $P_{\rm sat}$ detectors (\textit{LiteBIRD}-like), to fully understand the pick-up mechanism (possibly using the same readout of the flight module), and to understand its time-evolution and possible synchronization with the HWP rotation.

\acknowledgments    
We thank all \textit{LiteBIRD} collaborators for support and help. In particular Giovanni Signorelli, Joshua Montgomery and Samantha Lynn Stever for usuful comments and feedback on the manuscript. This work was supported by JSPS KAKENHI Grant Numbers 22K14054 and 18KK0083 and JSPS Core-to-Core Program, A. Advanced Research Networks. Kavli IPMU is supported by World Premier International Research Center Initiative (WPI), MEXT, Japan. \textit{LiteBIRD} (phase A) activities are supported by the following funding sources: ISAS/JAXA, MEXT, JSPS, KEK (Japan); CSA (Canada); CNES, CNRS, CEA (France); DFG (Germany); ASI, INFN, INAF (Italy); RCN (Norway); AEI (Spain); SNSA, SRC (Sweden); NASA, DOE (USA).

% References
\bibliography{main} % bibliography data in report.bib

\begin{thebibliography}{10}

\bibitem{simons_obs_2019}
{Ade}, P. et~al., ``{The Simons Observatory: science goals and forecasts},''
  {\em JCAP}~{\bf 2019},  056 (Feb. 2019).

\bibitem{bicep_2022}
{Ade}, P.~A.~R. et~al., ``{BICEP/Keck XV: The BICEP3 Cosmic Microwave
  Background Polarimeter and the First Three-year Data Set},'' {\em ApJ}~{\bf
  927},  77 (Mar. 2022).

\bibitem{litebird_ptep}
{LiteBIRD Collaboration}, {Allys}, E., et~al., ``{Probing Cosmic Inflation with
  the LiteBIRD Cosmic Microwave Background Polarization Survey},'' {\em arXiv
  e-prints} ,  arXiv:2202.02773 (Feb. 2022).

\bibitem{cmbs4_2022}
{Abazajian}, K. et~al., ``{CMB-S4: Forecasting Constraints on Primordial
  Gravitational Waves},'' {\em ApJ}~{\bf 926},  54 (Feb. 2022).

\bibitem{guth_1981}
{Guth}, A.~H., ``{Inflationary universe: A possible solution to the horizon and
  flatness problems},'' {\em Phys. Rev. D}~{\bf 23},  347--356 (Jan. 1981).

\bibitem{tristram_2022}
{Tristram}, M. et~al., ``{Improved limits on the tensor-to-scalar ratio using
  BICEP and P l a n c k data},'' {\em Phys. Rev. D}~{\bf 105},  083524 (Apr.
  2022).

\bibitem{irwin_2005}
{Irwin}, K.~D. and {Hilton}, G.~C., ``{Transition-Edge Sensors},'' in [{\em
  Cryogenic Particle Detection}{\nolinebreak\hspace{0.1em}]},  {Enss}, C., ed.,
   {\bf 99},  63 (2005).

\bibitem{barron_2018}
{Barron}, D. et~al., ``{Optimization study for the experimental configuration
  of CMB-S4},'' {\em JCAP}~{\bf 2018},  009 (Feb. 2018).

\bibitem{takakura_2019}
{Takakura}, S. et~al., ``{Measurements of Tropospheric Ice Clouds with a
  Ground-based CMB Polarization Experiment, POLARBEAR},'' {\em ApJ}~{\bf 870},
  102 (Jan. 2019).

\bibitem{puglisi_2022}
{Puglisi}, G. et~al., ``{Improved galactic foreground removal for B-mode
  detection with clustering methods},'' {\em MNRAS}~{\bf 511},  2052--2074
  (Apr. 2022).

\bibitem{ghigna_2020}
{Ghigna}, T., {Matsumura}, T., {Patanchon}, G., {Ishino}, H., and {Hazumi}, M.,
  ``{Requirements for future CMB satellite missions: photometric and band-pass
  response calibration},'' {\em JCAP}~{\bf 2020},  030 (Nov. 2020).

\bibitem{puglisi_2021}
{Puglisi}, G. et~al., ``{Simulating Calibration and Beam Systematics for a
  Future CMB Space Mission with the TOAST Package},'' {\em Research Notes of
  the American Astronomical Society}~{\bf 5},  137 (June 2021).

\bibitem{fabbian_2021}
{Fabbian}, G. and {Peloton}, J., ``{Simulating instrumental systematics of
  Cosmic Microwave Background experiments with s4cmb},'' {\em The Journal of
  Open Source Software}~{\bf 6},  3022 (Apr. 2021).

\bibitem{giardiello_2022}
{Giardiello}, S. et~al., ``{Detailed study of HWP non-idealities and their
  impact on future measurements of CMB polarization anisotropies from space},''
  {\em A\mbox{\&}A}~{\bf 658},  A15 (Feb. 2022).

\bibitem{krachmalnicoff_2022}
{Krachmalnicoff}, N. et~al., ``{In-flight polarization angle calibration for
  LiteBIRD: blind challenge and cosmological implications},'' {\em JCAP}~{\bf
  2022},  039 (Jan. 2022).

\bibitem{takakura_2017}
{Takakura}, S. et~al., ``{Performance of a continuously rotating half-wave
  plate on the POLARBEAR telescope},'' {\em JCAP}~{\bf 2017},  008 (May 2017).

\bibitem{sekimoto_2020}
Sekimoto, Y. et~al., ``{Concept design of low frequency telescope for CMB
  B-mode polarization satellite LiteBIRD},'' in [{\em Millimeter,
  Submillimeter, and Far-Infrared Detectors and Instrumentation for Astronomy
  X}{\nolinebreak\hspace{0.1em}]},  Zmuidzinas, J. and Gao, J.-R., eds.,  {\bf
  11453},  189 -- 209, International Society for Optics and Photonics, SPIE
  (2020).

\bibitem{sakurai_2020}
{Sakurai}, Y. et~al., ``{Breadboard model of the polarization modulator unit
  based on a continuously rotating half-wave plate for the low-frequency
  telescope of the LiteBIRD space mission},'' in [{\em Society of Photo-Optical
  Instrumentation Engineers (SPIE) Conference
  Series}{\nolinebreak\hspace{0.1em}]},  {\em Society of Photo-Optical
  Instrumentation Engineers (SPIE) Conference Series} {\bf 11453},  114534E
  (Dec. 2020).

\bibitem{hoang_2017}
{Thuong Hoang}, D. et~al., ``{Bandpass mismatch error for satellite CMB
  experiments I: estimating the spurious signal},'' {\em JCAP}~{\bf 2017},  015
  (Dec. 2017).

\bibitem{takaku_2022}
{Takaku}, R. et~al., ``{Evaluation of instrumental polarization induced by
  broadband achromatic half-wave plate with anti-reflective structures},'' in
  [{\em Society of Photo-Optical Instrumentation Engineers (SPIE) Conference
  Series}{\nolinebreak\hspace{0.1em}]},  {\em in preparation}  (2022).

\bibitem{hasebe_2022}
{Hasebe}, T. et~al., ``{Heat dissipation of rotation mechanism of polarization
  modulator unit for LiteBIRD low-frequency telescope},'' {\em in
  prepreparation}  (2022).

\bibitem{vavagiakis_2018}
{Vavagiakis}, E.~M. et~al., ``{Magnetic Sensitivity of AlMn TESes and Shielding
  Considerations for Next-Generation CMB Surveys},'' {\em Journal of Low
  Temperature Physics}~{\bf 193},  288--297 (Nov. 2018).

\bibitem{westbrook_2020}
Westbrook, B. et~al., ``{Detector fabrication development for the LiteBIRD
  satellite mission},'' in [{\em Space Telescopes and Instrumentation 2020:
  Optical, Infrared, and Millimeter Wave}{\nolinebreak\hspace{0.1em}]},
  Lystrup, M., Perrin, M.~D., Batalha, N., Siegler, N., and Tong, E.~C., eds.,
  {\bf 11443},  915 -- 936, International Society for Optics and Photonics,
  SPIE (2020).

\bibitem{westbrook_2022}
{Westbrook}, B. et~al., ``{Development of the Low Frequency Telescope Focal
  Plane Detector Arrays for LiteBIRD},'' {\em in preparation}  (2022).

\bibitem{niemack_2010}
{Niemack}, M.~D. et~al., ``{ACTPol: a polarization-sensitive receiver for the
  Atacama Cosmology Telescope},'' in [{\em Millimeter, Submillimeter, and
  Far-Infrared Detectors and Instrumentation for Astronomy
  V}{\nolinebreak\hspace{0.1em}]},  {Holland}, W.~S. and {Zmuidzinas}, J.,
  eds., {\em Society of Photo-Optical Instrumentation Engineers (SPIE)
  Conference Series} {\bf 7741},  77411S (July 2010).

\bibitem{thornton_2016}
{Thornton}, R.~J. et~al., ``{The Atacama Cosmology Telescope: The
  Polarization-sensitive ACTPol Instrument},'' {\em ApJ}~{\bf 227},  21 (Dec.
  2016).

\bibitem{inoue_2016}
{Inoue}, Y. et~al., ``{POLARBEAR-2: an instrument for CMB polarization
  measurements},'' in [{\em Millimeter, Submillimeter, and Far-Infrared
  Detectors and Instrumentation for Astronomy
  VIII}{\nolinebreak\hspace{0.1em}]},  {Holland}, W.~S. and {Zmuidzinas}, J.,
  eds., {\em Society of Photo-Optical Instrumentation Engineers (SPIE)
  Conference Series} {\bf 9914},  99141I (July 2016).

\bibitem{suzuki_phd}
{Suzuki}, A., {\em {Multichroic Bolometric Detector Architecture for Cosmic
  Microwave Background Polarimetry Experiments}}, PhD thesis, University of
  California, Berkeley (Jan. 2013).

\bibitem{jaehnig_2022}
{Jaehnig}, G. et~al., ``{LiteBIRD Low and Mid-frequency Telescope Detectors
  Design and Status},'' {\em in preparation}  (2022).

\bibitem{sugiyama_2022}
{Sugiyama}, S. et~al., ``{Vibration characteristics of a continuously rotating
  HWP using superconducting magnetic bearing and its potential impact on TES
  detectors and SQUIDs for CMB polarimetry},'' {\em in preparation}  (2022).

\bibitem{ghigna_jltp_2020}
{Ghigna}, T. et~al., ``{Design of a Testbed for the Study of System
  Interference in Space CMB Polarimetry},'' {\em Journal of Low Temperature
  Physics}~{\bf 199},  622--630 (Jan. 2020).

\bibitem{ghigna_phd}
{Ghigna}, T., {\em {Development of new generation receivers for experimental
  cosmology with the cosmic microwave background and systematic effect
  studies}}, PhD thesis, University of Oxford (2020).

\bibitem{dobbs_2012}
{Dobbs}, M.~A. et~al., ``{Frequency multiplexed superconducting quantum
  interference device readout of large bolometer arrays for cosmic microwave
  background measurements},'' {\em Review of Scientific Instruments}~{\bf 83},
  073113--073113--24 (July 2012).

\bibitem{montgomery_2022}
{Montgomery}, J. et~al., ``{Performance and characterization of the SPT-3G
  digital frequency-domain multiplexed readout system using an improved noise
  and crosstalk model},'' {\em Journal of Astronomical Telescopes, Instruments,
  and Systems}~{\bf 8},  014001 (Jan. 2022).

\bibitem{elleflot_2022}
{Elleflot}, T. et~al., ``{Low Noise Frequency Domain Multiplexing of TES
  Bolometers using Sub-kelvin SQUIDs},'' {\em arXiv e-prints} ,
  arXiv:2112.02425 (Dec. 2021).

\end{thebibliography}
\bibliographystyle{spiebib} % makes bibtex use spiebib.bst

\end{document}